\documentclass[twocolumn,amsmath,amssymb,prl]{revtex4}

\usepackage{amssymb}
\usepackage{amsmath}
\usepackage{graphicx,amssymb,mathtools,mathptmx}

\usepackage{bm,times}
\usepackage{xcolor}
\usepackage[%
colorlinks=true,
urlcolor=blue,
linkcolor=blue,
citecolor=blue
]{hyperref}
\usepackage{subfigure}
\usepackage[english]{babel}
\usepackage{color}
\usepackage{soul}
\usepackage{relsize}
\usepackage{scalerel}
\usepackage{appendix}
\usepackage{physics}
\usepackage{amsmath}
\usepackage{lipsum}

\begin{document}

\title{long-distance high-fidelity continuous-variable quantum key distribution with non-Gaussian operations: An exact closed form solution}

\author{Khatereh Jafari $^{1*}$, Mojtaba Golshani $^{2}$, Alireza Bahrampour $^{1}$}
\affiliation{
	$^1$Department of Physics, Sharif University of Technology, Tehran, Iran
	\\
	$^2$Faculty of Physics, Shahid Bahonar University of Kerman, Kerman, Iran
	\\
}

%
%




\vspace{-10mm}

\begin{abstract}
 In this paper, we derive a closed-form expression for the output state of a CV-QKD protocol in the presence of zero-photon catalysis (ZPC) and a quantum scissor (QS). Then, based on this closed-form solution, we use a direct search algorithm to find the appropriate values of input state and QS parameters, which considerably enhance the range and the fidelity of a CV-QKD protocol. In the special case of pure loss channel, the largest range of the protocol is only $6.5\%$ less than the fundamental limit of repeaterless quantum communication. In addition, examination of the protocol for different values of excess noise, reveals that there is a trade-off between range and fidelity, and a high value of fidelity can be obtained at the cost of a slight reduction in protocol range.
 \\
 \\
 Keywords: Continuous-Variable Quantum Key Distribution, Secure Key Rate, Zero Photon Catalysis, Quantum Scissor, Long-range, High-Fidelity
 \vspace{10mm}
 \\$^*$jafari831@gmail.com\\
\end{abstract}

\maketitle

\section{Introduction}
Unlike classical cryptography, whose security is due to the high complexity of the mathematical problem, quantum key distribution (QKD) is a method that shares an unconditionally secure key between two users (traditionally called Alice and Bob) based on fundamental laws of quantum physics\cite{wolf2021quantum,pirandola2020advances,gisin2002quantum}. In the continuous variable QKD (CV-QKD), the information is encoded on the continuous characteristics of the light, e.g. its quadratures\cite{ralph1999continuous,PhysRevLett.88.057902,grosshans2003quantum,weedbrook2012gaussian,braunstein2005quantum}.
Compared to its discrete counterpart\cite{Bennett-2014,PhysRevLett.67.661,bennett1992experimental}, CV-QKD has some advantages, namely compatibility with conventional technology, low cost, and high rate\cite{hirano2003quantum,yonezawa2007experimental}. Nevertheless, this protocol has drawbacks such as short range and sensitivity to the excess noise\cite{ma2019long}. To dispel these problems, different setups including quantum scissor (QS), photon addition, photon subtraction, and photon catalysis are proposed\cite{PhysRevA.102.012608,guo2019continuous,ye2019improvement,blandino2012improving}.

QS is a non-deterministic operation that reduces the dimension of Hilbert space into two dimensions, and provides an arbitrary linear combination of vacuum and single photon states\cite{pegg1998optical}. Based on the transmittance coefficients of scissor’s beam splitters, QS can act as a heralded noiseless linear amplifier for small amplitude input signals, and consequently can boost the performance of CV-QKD protocols at long distances\cite{Ralph2009NondeterministicNL,ghalaii2020long,ghalaii2020discrete}.

Photon Catalysis is a feasible approach to improve the performance of CV-QKD protocols\cite{guo2019continuous,ye2019improvement,bartley2015directly}.
Zero photon catalysis (ZPC) is a special type of quantum catalysis that reduces the weight of higher-order Fock states and can be treated as a heralded noiseless attenuator\cite{bartley2015directly}. This Gaussian operation, by reducing the modulation variance of the input signal, is able to increase the secure transmission distance of the CV-QKD protocols.

Quantum states become disturbed when passing through a quantum channel. In quantum communication, the main goal is to increase the secrete key rate and the secure transmission distance of the protocol, but it is also important to reduce errors caused by the channel. In fact, reducing the errors and its effect on the corruption of the transmitted state is important in the error correction process (post-processing stage)~\cite{matsumoto2021survey,salim2020enhancing,bouchard2022quantum}. The fidelity is a measure of the closeness of Alice and Bob states, and hence can be used to quantify the corruption of states during the quantum channel~\cite{wolf2021quantum}. In addition, some QKD protocols such as measurement-device-independent CV-QKD (MDI-CV-QKD) protocols are based on the teleportation of quantum states in which the fidelity plays a vital role~\cite{PhysRevLett.108.130503,PhysRevA.89.052301,PhysRevA.87.022308,HighRate397NatPhotonics}. Therefore, it will be important to find the optimal parameters that can enhance the fidelity without significant reduction of the protocol range.

Recently, we offered a CV-QKD protocol equipped with both ZPC and QS, and demonstrated that this suggested scheme outstandingly enhances the fidelity  even at long distances~\cite{jafari2023high}. 
However, in that work, due to the structure of the protocol, it was not possible to obtain a closed-form solution for the output state of the quantum channel. Therefore,
due to the time-consuming numerical calculations, the optimal values of transmittance coefficients of QS's beam splitters, to achieve the maximum secure range, have not been obtained. 
In this paper, we are looking to optimize the parameters to obtain the highest achievable values of the protocol range and fidelity.
In order to realize this, some parts of the previous protocol have been changed. In Ref.~\cite{jafari2023high}, we used a modified QS that truncates the Hilbert space to third-order Fock states\cite{jafari2022discrete}. However, in this research, the standard QS is adopted.
This modification considerably simplifies both the experimental setup (because of the simpler sources and detectors of the standard QS) and computational complexity (optimization of one beam splitter instead of two).
Moreover, here, we assume Alice uses Gaussian modulation in contrast to the previous work that discrete (quadrature phase-shift keying (QPSK)) modulation was adopted\cite{leverrier2009unconditional}. In the entanglement-assisted protocol, this means Alice uses a two-mode squeezed vacuum (TMSV) state and performs heterodyne measurement on her mode\cite{djordjevic2019physical}. In this case, exact closed-form expression can be obtained for the output state of the protocol, which significantly reduces the required computational time of the optimization process. Moreover, As we will see in the next section, performing of the ZPC operation on one mode of the TMSV state (in contrast to the QPSK state) is equivalent to reducing the modulation variance of that state. Therefore, it is possible to further simplify the protocol and eliminate the ZPC operation, and as an alternative optimize the modulation variance of the TMSV state.

In this paper, we derive an exact expression for the output state of a protocol in the presence of standard QS at the Bob side. Then, based on this result, the optimal modulation variance and transmission coefficient of QS are numerically calculated to achieve the maximum secure transmission distance. As will be seen, this optimization allows us to find the parameters that considerably  boost the secure range of the protocol, as well as, leads to a high fidelity between input and output states.
These findings 
highlight the importance of the optimization in CV-QKD protocols.

This research is organized in four sections. The protocol description and theoretical consideration is described in Section 2. In section 3, we present the optimization of the protocol and discuss about its numerical results. Finally, conclusion is presented in section 4.

\section{Protocol Description and Theoretical Consideration}
\label{theory}

\begin{figure}[t!]
	\centering
	\includegraphics[width=1.0\columnwidth]{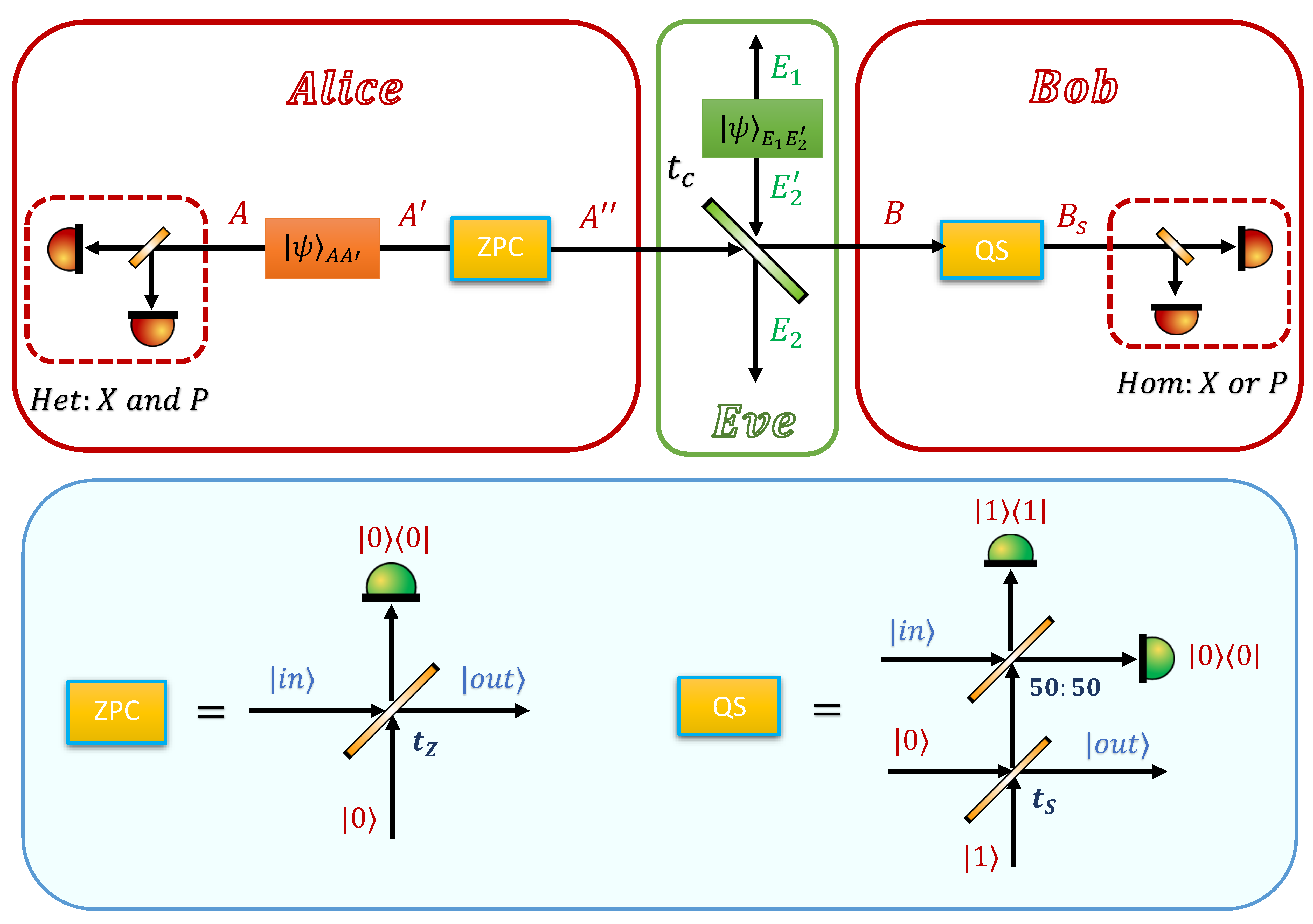}
	\caption{
		\label{fig1}%
		Entanglement-based scheme of the protocol.}
\end{figure}

\begin{figure}[t!]
	\centering
	\includegraphics[width=1.0\columnwidth]{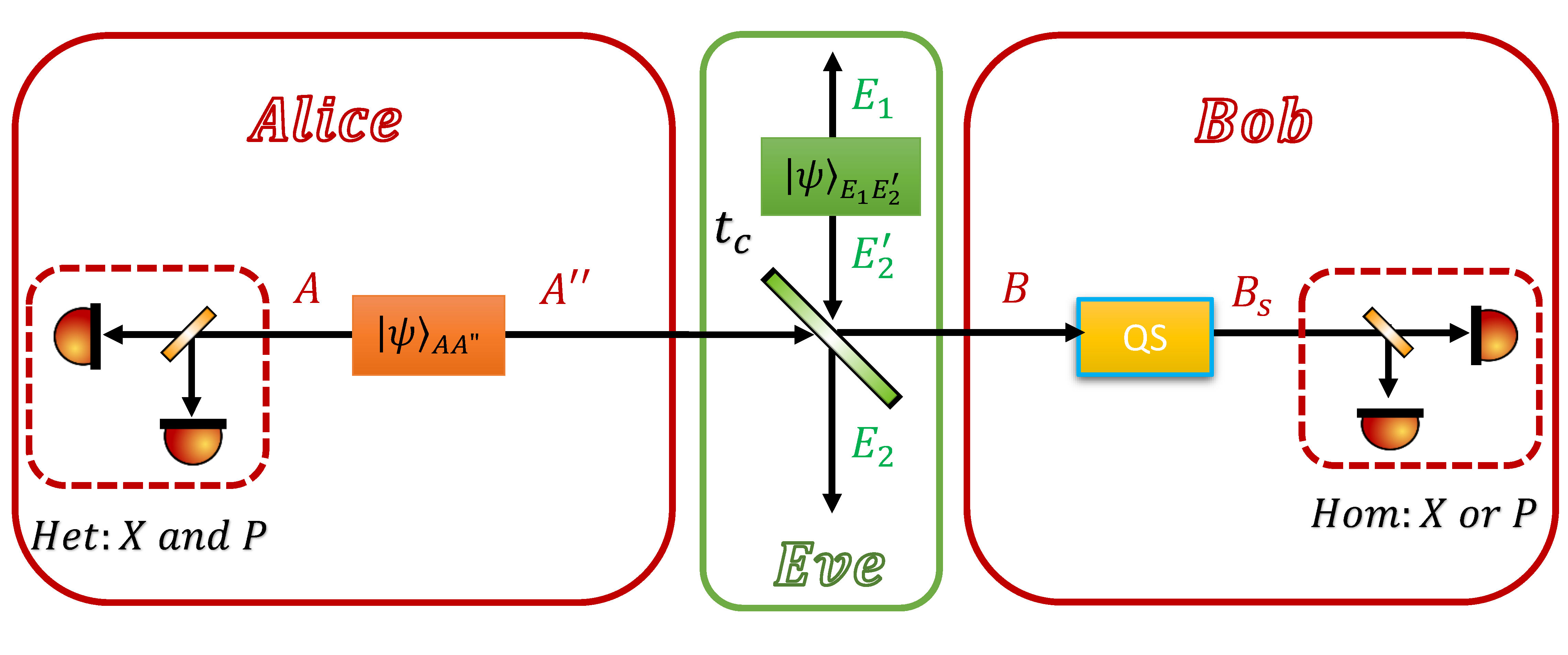}
	\caption{
		\label{fig2}%
		Simplified version of the protocol.}
\end{figure}

The schematic of the protocol is illustrated in Fig.~\ref{fig1}, which uses a ZPC at Alice’s side and a QS at Bob’s side. In general, the Alice state can be written in the number state basis as,
$
\ket{\psi}_{A A'}=\sum_{n_A,n_A'}{\Gamma(n_A,n_A')\ket{n_A}_{A}\ket{n_A'}_{A'}},
$
where $\Gamma(n_A,n_A')$ satisfies the normalization condition,
$
\sum_{n_A,n_A'}{\left|\Gamma(n_A,n_A')\right|^2}=1.
$
Alice performs a ZPC operation on mode $A'$, which converts $\Gamma(n_A,n_A')$ to
$
\frac{t_z^{n_A'}}{\sqrt{P_z}}\Gamma(n_A,n_A'),
$
where $t_z$ denotes the transmission coefficient of the beam splitter and the normalization factor 
$
P_z=\sum_{n_A,n_A'}{\left|t_z^{n_A'}\Gamma(n_A,n_A')\right|^2}
$
is the success probability of ZPC operation\cite{jafari2023high}. Since $\left| t_z\right|<1$, ZPC operation acts as an attenuator, reduces the contribution of higher-order number states by the factor $\left|t_z\right|^{2n_A'}$ and therefore degrades the impact of channel loss on the input state. Conversely, we use a QS at the end of the channel, which acts as a noiseless amplifier, compensates this weight reduction, and upgrades the performance of the protocol.
In this paper, we assume Alice uses a TMSV state
$
\ket{\psi}_{A A'}=\sqrt{1-\lambda_A^2}\sum_{n_A=0}^{\infty}{\lambda_A^{n_A}\ket{n_A}_{A}\ket{n_A}_{A'}},
$
which is a special case with 
$\Gamma(n_A,n_A')=\sqrt{1-\lambda_A^2}\,\lambda_A^{n_A} \delta_{n_A,n_A'}$
($\delta_{n_A,n_A'}$ is the Kronecker delta).
ZPC operation on mode $A'$ changes this state to the state
$
\ket{\psi}_{A A''}=\sqrt{\frac{1-\lambda_A^2}{P_z}}\sum_{n_A=0}^{\infty}{t_z^{n_A}\lambda_A^{n_A}\ket{n_A}_{A}\ket{n_A}_{A''}},
$
with $P_z=\frac{1-\lambda_A^2}{1-t_z^2\lambda_A^2}$. For TMSV state, ZPC operation exclusively reduces $\lambda_A$ by the factor $t_z$, i.e., diminishes the quadrature variance from $\frac{1+\lambda_A^2}{1-\lambda_A^2}$ to $\frac{1+t_z^2\lambda_A^2}{1-t_z^2\lambda_A^2}$ \cite{djordjevic2019physical}.
Accordingly, for this special input state, one can remove the ZPC operation (i.e., set $t_z=1$) and apply the optimization process to $\lambda_A$ (see Fig.~\ref{fig2}). 
It is important to note that this conclusion is not true for a general input state (for example, QPSK modulation scheme of Ref.~\cite{jafari2023high}). Therefore, in general, both ZPC operation and QS are required to noiselessly diminish the loss of quantum channel and improve the performance of the protocol~\cite{mivcuda2012noiseless}.
Alice sends mode $A''$ through the quantum channel and performs heterodyne measurement on her mode $A$. In this paper, we assume that two parties use a thermal-loss channel with transitivity $t_c^2$ and an excess noise $\varepsilon$. This kind of channel can be modeled with a beam splitter with the transmission coefficient $t_c=10^{-0.01 L}$, where $L$ is the channel length (in km unit)~\cite{ghalaii2020discrete,ghalaii2020long}. This assumption corresponds to the case of a Gaussian attack enabled by an entangling cloner, in which one input of the beam splitter is Alice's mode $A''$ and the other is mode $E'_2$ of Eve's TMSV state
$
\ket{\psi}_{E_1 E'_2}=\sqrt{1-\lambda_E^2}\sum_{n_E=0}^{\infty}{\lambda_E^{n_E}\ket{n_E}_{E_1}\ket{n_E}_{E'_2}}.
$
Here,  $\lambda_E$ characterizes the excess noise of the channel, $\varepsilon=\frac{2\lambda_E^2}{1-\lambda_E^2}$\cite{jafari2022discrete}.
In this case, the equivalent excess noise at the transmitter side (input of the channel) is given by 
$\varepsilon_{tm}=(1-t_c^2)\varepsilon/t_c^2$, while 
$\varepsilon_{rec}=(1-t_c^2)\varepsilon$ represents the amount of excess noise at
the receiver side (output of the channel)~\cite{ghalaii2020discrete,ghalaii2020long}. Here, same as Refs.~\cite{jafari2022discrete,jafari2023high} we use 
$\varepsilon$, and investigate this proposed protocol for different values of  $\varepsilon$. 
It is important to note that  $\varepsilon_{rec}$ is always less than $\varepsilon$ and approaches it in the limit of large values of propagation distances $L$, but, $\varepsilon_{tm}$ is larger than $\varepsilon$ for $L\gtrsim 15\mathrm{km}$, and gets larger by increasing the propagation distances $L$. As we will see later, in the range of values considered in this study (see Table.~\ref{tablenumericalvalues}), $\varepsilon$ is very close to $\varepsilon_{rec}$, and therefore, the excess noise $\varepsilon$ can be considered as the noise at the output of the channel. 
\\
Our protocol uses a QS that is a non-Gaussian operation. Therefore, Gaussian attack is not necessarily an optimal attack for this protocol. Finding optimal eavesdropper strategy may be very challenging and is not the intention of current research. Therefore, we restrict the eavesdropper to a Gaussian attack.  We will talk about this issue again later in the paper.

The states
$\ket{\psi}_{A A''}$ and $\ket{\psi}_{E_1 E'_2} $ after mixing in the beam splitter are converted to\cite{jafari2023high}
\begin{eqnarray}
\label{eqOutputstate}
\ket{\psi}_{\scaleto{A B E_1 E_2}{5.5pt}}=\mathlarger{\sum}_{\substack{n_A,n_B \\ n_E}}{\Theta_{n_A,n_B}^{n_E} \scaleto{\ket{n_A}_{A}\ket{n_B}_{B}\ket{n_E}_{E_1}\ket{n_A+n_E-n_B}_{E_2}}{10.5pt}}.
\end{eqnarray}
where,
\small
\begin{eqnarray*}
\label{eqTheta1}
\begin{split}
&\Theta_{n_A,n_B}^{n_E}=\sqrt{(1-\lambda_A^2)(1-\lambda_E^2)}\lambda_A^{n_A}\lambda_E^{n_E}(-t_c)^{n_E-n_B} r_c^{n_A+n_B}\binom{n_E}{n_B}
\\
&\times \sqrt{\frac{n_B!(n_A+n_E-n_B)!}{n_A! n_E!}} {}_2\mathlarger{F}_1\left(-n_A,-n_B;1+n_E-n_B;-\frac{t_c^2}{r_c^2} \right),
\end{split}
\end{eqnarray*}
\normalsize
for $0 \le n_B \le n_E$, and
\small
\begin{eqnarray*}
\label{eqTheta2}
\begin{split}
&\Theta_{n_A,n_B}^{n_E}=\sqrt{(1-\lambda_A^2)(1-\lambda_E^2)}\lambda_A^{n_A}\lambda_E^{n_E} t_c^{n_B-n_E} r_c^{n_A-n_B+2n_E}\binom{n_A}{n_B-n_E}
\\
&\times \sqrt{\frac{n_B!(n_A+n_E-n_B)!}{n_A! n_E!}} {}_2\mathlarger{F}_1\left(n_B-n_A-n_E,-n_E;1+n_B-n_E;-\frac{t_c^2}{r_c^2} \right),
\end{split}
\end{eqnarray*}
\normalsize
for $n_E <n_B \le n_A+n_E$.
Here, $r_c=\sqrt{1-t_c^2}$ is the reflection coefficient of the beam splitter and ${}_2F_1(a,b;c;z)$ is the Gaussian or ordinary hypergeometric function~\cite{arfken2011mathematical}.

The noise and loss of the channel limit the secure range of the protocol. To overcome this limitation, we assume Bob is equipped with a QS. QS is a heralded noiseless linear amplifier that can compensate the loss of quantum channel and boosts the secure transmission distance of the protocol\cite{jafari2022discrete}. Standard QS cuts off all Fock states with two or more photons and prepares a superposition of zero and one photon states~\cite{pegg1998optical}. Setting $n_B=0$ (zero photon) and $n_B=1$ (one photon) in $\Theta_{n_A,n_B}^{n_E}$, after straightforward calculations, one can find,
\begin{eqnarray}
\label{eqOutputstateScissor}
\begin{split}
\ket{\psi}_{\scaleto{A B_s E_1 E_2}{5.5pt}}=&\mathlarger{\sum}_{n_A,n_E}{V_{n_A}^{n_E} \scaleto{\ket{n_A}_{A}\ket{0}_{B_s}\ket{n_E}_{E_1}\ket{n_A+n_E}_{E_2}}{10.5pt}}
\\
+&\mathlarger{\sum}_{n_A,n_E}{W_{n_A}^{n_E} \scaleto{\ket{n_A}_{A}\ket{1}_{B_s}\ket{n_E}_{E_1}\ket{n_A+n_E-1}_{E_2}}{10.5pt}}.
\end{split}
\end{eqnarray}
where
\small
\begin{eqnarray*}
\label{eqW}
V_{n_A}^{n_E}=t_s(-t_c)^{n_E}r_c^{n_A}\lambda_A^{n_A}\lambda_E^{n_E}\sqrt{\frac{(1-\lambda_A^2)(1-\lambda_E^2)(n_A+n_E)!}{2 n_A! n_E! P_Q}},
\end{eqnarray*}
\normalsize
and
\small
\begin{eqnarray*}
\label{eqW}
\begin{split}
W_{n_A}^{n_E}=&gt_s(-t_c)^{n_E-1}r_c^{n_A-1}(n_E r_c^2-n_At_c^2) \lambda_A^{n_A}\lambda_E^{n_E}
\\
&\times \sqrt{\frac{(1-\lambda_A^2)(1-\lambda_E^2)(n_A+n_E-1)!}{2 n_A! n_E! P_Q}}.
\end{split}
\end{eqnarray*}
\normalsize
Here, $g=\sqrt{\frac{1-t_s^2}{t_s^2}}$ and $t_s$ are the amplification gain and transmission coefficient of QS, respectively~\cite{Ralph2009NondeterministicNL}.
Moreover, the normalization factor
\small
\begin{eqnarray}
\label{eqSuccessScissor}
P_Q=\frac{(1-\lambda_A^2)(1-\lambda_E^2)[t_s^2(1-\lambda_A^2)(1-\lambda_E^2)+t_c^2\lambda_A^2+(r_c^2-\lambda_A^2)\lambda_E^2]}{2(1-r_c^2\lambda_A^2-t_c^2\lambda_E^2)^2}
\end{eqnarray}
\normalsize
is the success probability of QS, i.e., the probability that its detectors count zero and one photon at horizontal and vertical output arms, respectively (see the right bottom panel of Fig.~\ref{fig1}).

The Alice-Bob state $\rho_{AB_s}$ is the partial trance of Eq.~(\ref{eqOutputstateScissor}) over eavesdropper modes $E_1$ and $E_2$, and is given by
\begin{eqnarray}
\label{eqRhoABs}
\begin{split}
\rho_{AB_s}=&\mathlarger{\sum}_{n_A=0}^{\infty}\left(\sigma^{00}_{n_A}\ket{n_A,0}\bra{n_A,0}+\sigma^{11}_{n_A}\ket{n_A,1}\bra{n_A,1} \right. \\
& \left. \sigma^{01}_{n_A}\ket{n_A,0}\bra{n_A+1,1}+\sigma^{10}_{n_A}\ket{n_A+1,1}\bra{n_A,0} \right),
\end{split}
\end{eqnarray}
where
\small
\begin{eqnarray*}
\label{eqRhoABs00}
\sigma^{00}_{n_A}=\frac{t_s^2(1-\lambda_A^2)(1-\lambda_E^2)(r_c\lambda_A)^{2n_A}}{2P_Q(1-t_c^2\lambda_E^2)^{n_A+1}},
\end{eqnarray*}
\begin{eqnarray*}
\label{eqRhoABs11}
\sigma^{11}_{n_A}=\frac{g^2t_s^2(1-\lambda_A^2)(1-\lambda_E^2)(r_c\lambda_A)^{2n_A}\left[ n_At_c^2(1-\lambda_E^2)^2+r_c^4\lambda_E^2 \right]}{2P_Q r_c^2(1-t_c^2\lambda_E^2)^{n_A+2}},
\end{eqnarray*}
\begin{eqnarray*}
\label{eqRhoABs01}
\sigma^{01}_{n_A}=\sigma^{10}_{n_A}=\frac{gt_s^2t_c\sqrt{n_A+1}(1-\lambda_A^2)(1-\lambda_E^2)^2 r_c^{2n_A}\lambda_A^{2n_A+1}}{2P_Q (1-t_c^2\lambda_E^2)^{n_A+2}}.
\end{eqnarray*}
\normalsize
Eq.~(\ref{eqRhoABs}) gives the analytical expression for output state of the protocol. Using this equation, the Alice $\rho_A$ and Bob $\rho_{B_s}$ states are given by
\begin{eqnarray}
\label{eqRhoA}
\rho_{A}=\mathlarger{\sum}_{n_A=0}^{\infty}{\gamma^{(A)}_{n_A}\ket{n_A}\bra{n_A} },
\end{eqnarray}
\begin{eqnarray}
\label{eqRhoBs}
\rho_{B_s}=\gamma^{(B)}_0\ket{0}\bra{0}+\gamma^{(B)}_1\ket{1}\bra{1},
\end{eqnarray}
where
\small
\begin{eqnarray*}
\label{eqRhoAnA}
\begin{split}
\gamma^{(A)}_{n_A}&=\frac{(1-\lambda_A^2)(1-\lambda_E^2)(r_c\lambda_A)^{2n_A}}{2P_Q r_c^2(1-t_c^2\lambda_E^2)^{n_A+2}} \\
&\times \left[ n_A g^2 t_s^2 t_c^2(1-\lambda_E^2)^2+r_c^4\lambda_E^2 +r_c^2 t_s^2(1-\lambda_E^2) \right],
\end{split}
\end{eqnarray*}
\begin{eqnarray*}
\label{eqRhoB0}
\gamma^{(B)}_{0}=\frac{t_s^2(1-\lambda_A^2)(1-\lambda_E^2)}{2P_Q(1-r_c^2\lambda_A^2-t_c^2\lambda_E^2)},
\end{eqnarray*}
\begin{eqnarray*}
\label{eqRhoB1}
\gamma^{(B)}_{1}=\frac{g^2 t_s^2 (1-\lambda_A^2)(1-\lambda_E^2)[ t_c^2\lambda_A^2+\lambda_E^2(r_c^2-\lambda_A^2)]}{2P_Q(1-r_c^2\lambda_A^2-t_c^2\lambda_E^2)^2}.
\end{eqnarray*}
\normalsize
Here, same as~\cite{ghalaii2020long}, we can investigate the non-Gaussian behavior of our protocol. 
To this end, we focus on the distribution of homodyne measurement results on one of Bob’s quadratures, for example
$\hat{x}_{B_s}=\frac{\hat{a}_{B_s}+\hat{a}^{\dag}_{B_s}}{2}$.
The probability distribution of obtaining $x$ after measuring $\hat{x}_{B_s}$ is given by~\cite{jafari2022discrete} 
\begin{eqnarray}
\label{eqfxB}
f(x)=\bra{x}\rho_{B_s}\ket{x}=\sqrt{\frac{2}{\pi}}e^{-2x^2}\left( \gamma^{(B)}_{0} +4 \gamma^{(B)}_{1} x^2 \right),
\end{eqnarray}
where $\hat{x}_{B_s}\ket{x}=x\ket{x}$.
Accordingly, the Gaussian and non-Gaussian terms of $f(x)$ are proportional to $\gamma^{(B)}_{0}$ and $\gamma^{(B)}_{1}$, respectively. Therefore, the output state is approximately Gaussian as long as 
$\gamma^{(B)}_{1}/\gamma^{(B)}_{0} \ll 1$. Considering the fact that $\gamma^{(B)}_{1}$ decreases by decreasing of $t_c$ (i.e., increasing of propagation length $L$) and takes its largest value at $t_c=1$, one can simply show that the output state is close to the Gaussian state if $g^2\lambda_A^2 \ll 1$. If this condition is met, the Gaussian attack can be considered as the optimal attack.

To demonstrate the benefit of optimized input state and QS, we consider the reverse reconciliation scheme, in which Alice corrects her raw key in order to match Bob's key. Therefore, under collective attack, the secret key rate is given by
$K=P_Q\left( \beta S[\rho_A] - S[\rho_{AB_s}] \right)$, where $\beta$ is the reconciliation efficiency and $S[\rho]=-tr\left( \rho \log_2 \rho \right)$ is the Von Neumann entropy~\cite{jafari2022discrete}.
In this equation, $P_Q$ reduces the value of the key rate because the information is calculated only when QS is successful.
This is why QS diminishes the secure key rate at lower values of the propagation distance\cite{ghalaii2020discrete,ghalaii2020long}. However, at larger values of the propagation distance where the signal's strength gets weaker, QS acts as a noiseless amplifier, distills the entanglement, and improves the secure range of the protocol.
Although the secret key rate and the secure propagation distance are the key parameters, however,  fidelity is also an important parameter and high fidelity can simplify the post-processing stage of the QKD protocol~\cite{wolf2021quantum}. Therefore, in addition to the key rate, we will also discuss about the fidelity,
$
F=\left( tr \left( \sqrt{\sqrt{\rho_{B_s}} \rho_{A} \sqrt{\rho_{B_s}} } \right) \right)^2,
$
which is a popular measure of the closeness of the Alice $A$ and Bob $B_s$ modes~\cite{wolf2021quantum}. Using equations~(\ref{eqRhoA}) and (\ref{eqRhoBs}), the fidelity can be written as
\begin{eqnarray}
\label{eqFidelity}
F=\left( \sqrt{\gamma^{(A)}_{0}\gamma^{(B)}_{0}}+\sqrt{\gamma^{(A)}_{1}\gamma^{(B)}_{1}} \right)^2.
\end{eqnarray}
As we will see in the next section, optimization of input state (i.e., $\lambda_A$) and QS (i.e., $t_s$) results in long-distance high-fidelity QKD.

\section{Numerical results and Discussions}
\label{numericalresults}
Our goal is to find the optimal value of $\lambda_A$ and $t_s$ that maximize the protocol range.
TMSV state with $\lambda_A=1$ corresponds to maximally entangled (Einstein-Podolsky-Rosen) state, which is unnormalizable and unphysical\cite{braunstein2005quantum}. In fact, achieving values of $\lambda_A$ close to unity is practically difficult. 
In addition, QS acts as an amplifier ($g>1$) as long as $t_s^2<0.5$. 
Therefore, we restrict the search domain to $\lambda_A \in (0,0.9)$ and $t_s \in (0,0.7)$.
Exact equation for the output state of the channel allows us to use a direct search algorithm for this two-parameter optimization problem. To this end, we divide the intervals $0<\lambda_A \le 0.9$ and $0<t_s \le 0.7$ into 900 and 700 partitions, respectively, i.e. the step size $\Delta \lambda_A=\Delta t_s=0.001$. Then, for each value of $\lambda_A$ and $t_s$, we use the bisection method to find the longest propagation distance, i.e., protocol range $R$, that has a key rate greater than $K_{min}$. In this paper, we set the minimum acceptable secrete key rate (bits per pulse) as $K_{min}=10^{-6}$, while in literature smaller values of $10^{-8}$ and $10^{-10}$ are also considered~\cite{ghalaii2020long,ghalaii2020discrete,winnel2021overcoming}.
Finally, by comparing the protocol range for all $\lambda_A$ and $t_s$ values, the optimal value of the protocol range can be achieved.

\begin{figure}[t!]
	\centering
	\includegraphics[width=.99\columnwidth]{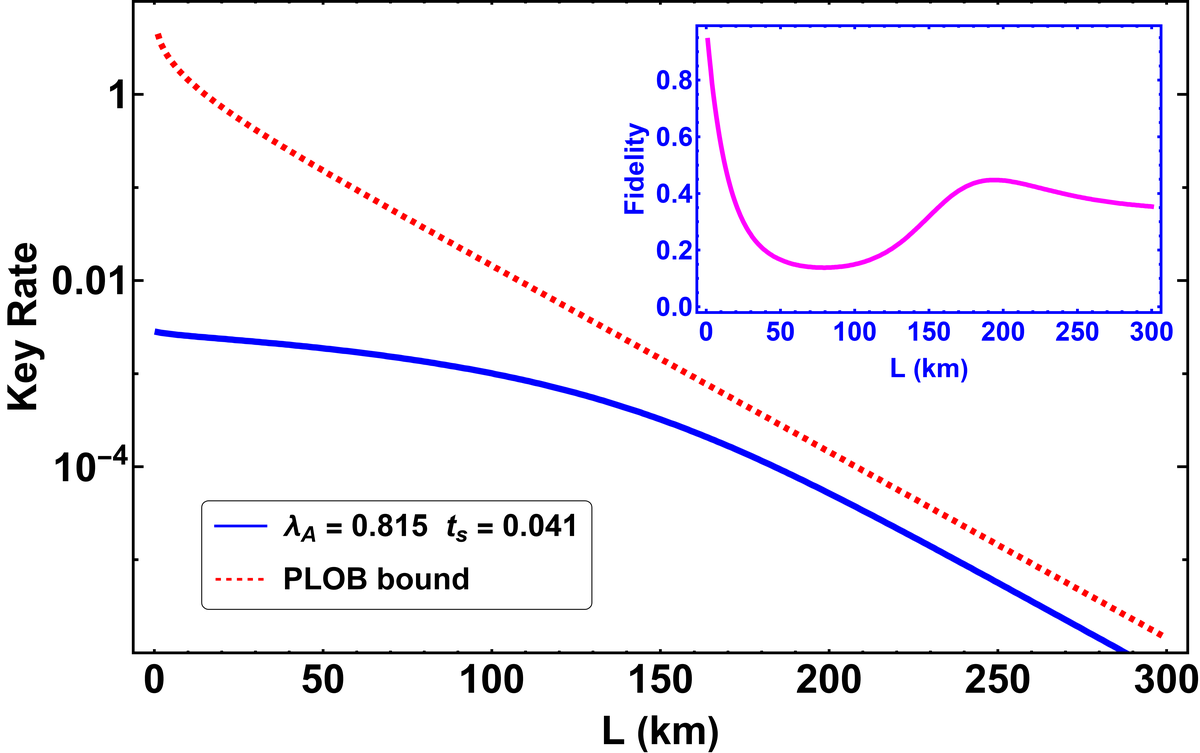}
	\caption{
		\label{figZeroNoiseMaxRangeKeyandFidelity}
		(Color online) Secret key rate (per pulse) versus transmission distance for $\lambda_A=0.815$, $t_s=0.041$ and $\epsilon \simeq 0$. For comparison, the PLOB bound is also shown. The inset shows fidelity versus transmission distance. }
\end{figure}
\begin{figure}
	\centering
	\includegraphics[width=0.99\columnwidth]{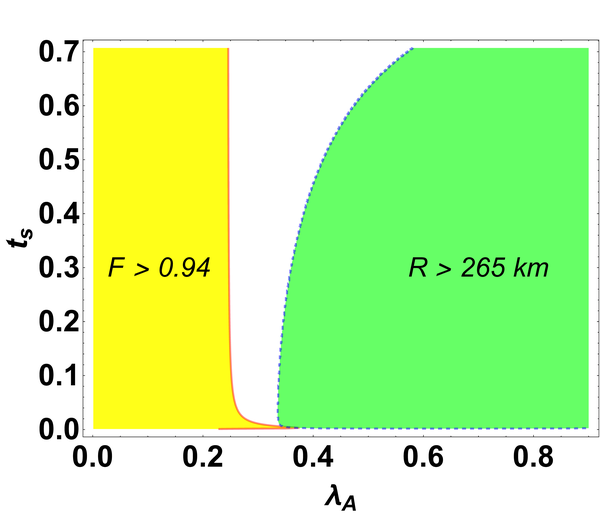}
	\caption{
		\label{figZeroNoiseContourRangeFidelity}
		(Color online) The green area shows the regions with the range larger than $265km$. The yellow area shows the regions with the fidelity greater than $0.94$ ($\epsilon \simeq 0$). }
\end{figure}

We first consider a pure loss (noiseless) channel. Therefore, we set $\varepsilon=0$ ($\lambda_E \to 0$). 
Our results show that, in this case, the maximum achievable protocol range is about $R=288 km$, which takes place at $\lambda_A=0.815$, at $t_s=0.041$.
Fig.~\ref{figZeroNoiseMaxRangeKeyandFidelity} shows the secret key rate versus transmission distance. For comparison, the PLOB bound (named after. Pirandola, Laurenza, Ottaviani, and Banchi) is also shown~\cite{pirandola2017fundamental}. The PLOB bound $K_{PLOB}=-log_2(1-t_c^2)$ is the fundamental limit of repeaterless quantum communication and reaches the value of $K_{min}=10^{-6}$ at $L \simeq 308 km$. This indicates that the maximum achievable range of our optimized protocol is only $6.5\%$ less than the PLOB range.
It is important to note that, QS reduces the value of the key rate at short distances because the information is calculated for the post-selected data only when QS is successful. However, at large distances, QS acts as an amplifier and boosts the protocol range~\cite{ghalaii2020long,ghalaii2020discrete}.

The inset of Fig.~\ref{figZeroNoiseMaxRangeKeyandFidelity} shows the fidelity $F$ of Eq.~\ref{eqFidelity} for $\lambda_A=0.815$ and $t_s=0.041$. According to this figure, the fidelity is about $0.36$  at the protocol range ($R=288km$), which is a small value. Therefore, we search for other values of $\lambda_A$ and $t_s$ with high range $R$ that have considerable fidelity $F$ at that range. Hereafter, the term 'fidelity' ($F$) would be used for the value of fidelity at the protocol range, i.e. at the propagation distance $L=R$.
The green and yellow areas of Fig.~\ref{figZeroNoiseContourRangeFidelity} show the regions in $\lambda_A-t_s$ plane with $R>265km$ and $F>0.94$, respectively.  
From this figure, one can see that the protocol has high range for larger values of $\lambda_A$ and smaller values of $t_s$. This means, when the input state has a large number of photons (large value of $\lambda_A$), or QS has a high amplification gain (small value of $t_s$), in the absence of noise, the output state is less altered by the loss, and therefore, the range of protocol increases. In addition, at smaller values of $\lambda_A$, in which the weight of higher-order number states of Alice mode are small, the fidelity between Alice (Eq.~\ref{eqRhoA}) and Bob (Eq.~\ref{eqRhoBs}) states increases.

\begin{figure}
	\centering
	\includegraphics[width=0.99\columnwidth]{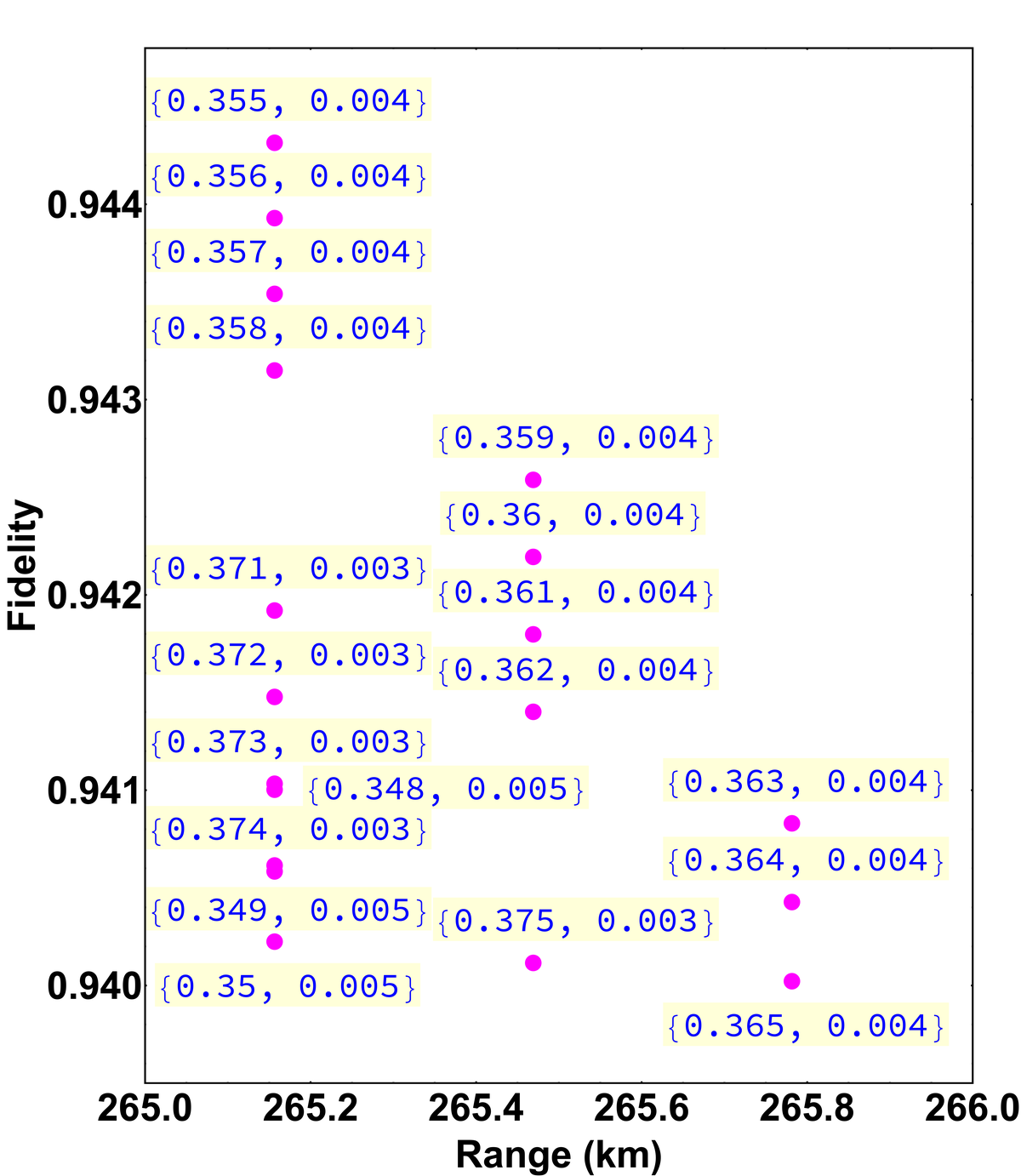}
	\caption{
		\label{figZeroNoiseRangeFidelityPoints}
		(Color online) The $(\lambda_A,t_s)$ points with a range larger than $265km$ and a fidelity greater than $0.94$ ($\epsilon \simeq 0$). }
\end{figure}
\begin{figure}
	\centering
	\includegraphics[width=0.99\columnwidth]{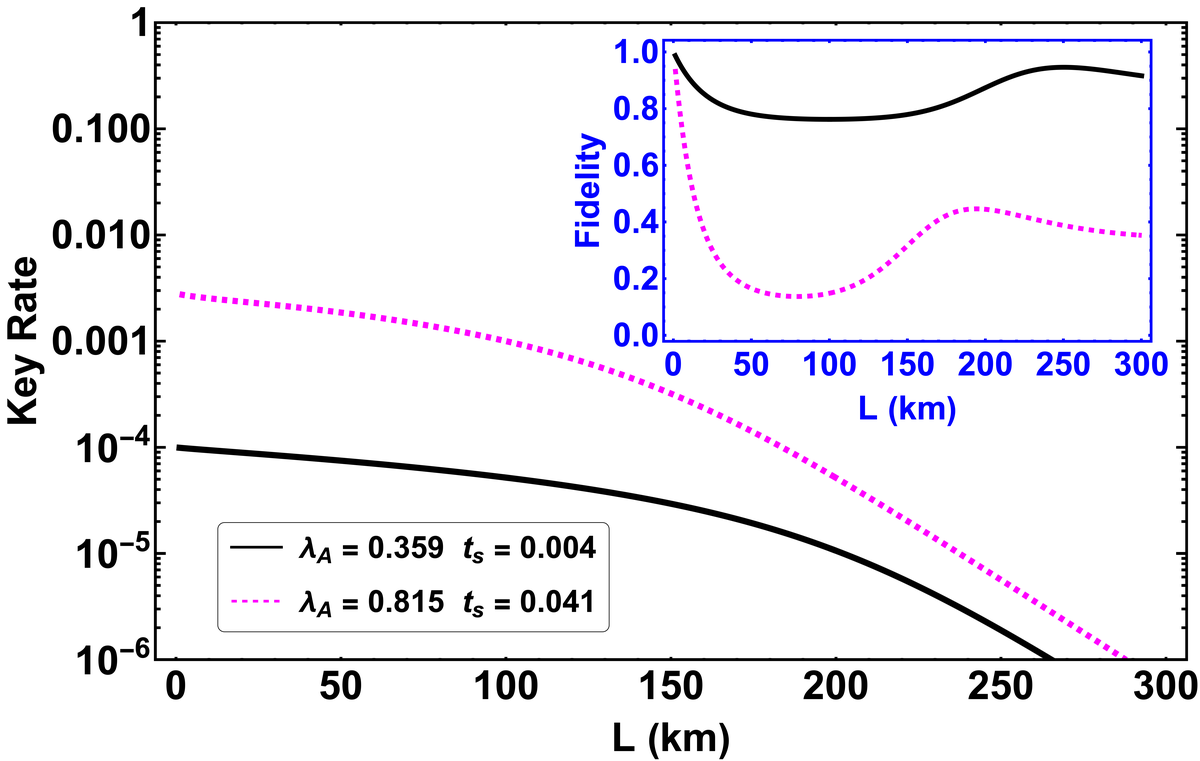}
	\caption{
		\label{figZeroNoiseMaxRangeandHighFidelityKeyandFidelity}
		(Color online) Secret key rate (per pulse) versus transmission distance for $\lambda_A=0.815$, $t_s=0.041$ and $\lambda_A=0.359$, $t_s=0.004$, for the case with $\epsilon\simeq 0$. The inset compares fidelities. }
\end{figure}
\begin{figure}
	\centering
	\includegraphics[width=0.99\columnwidth]{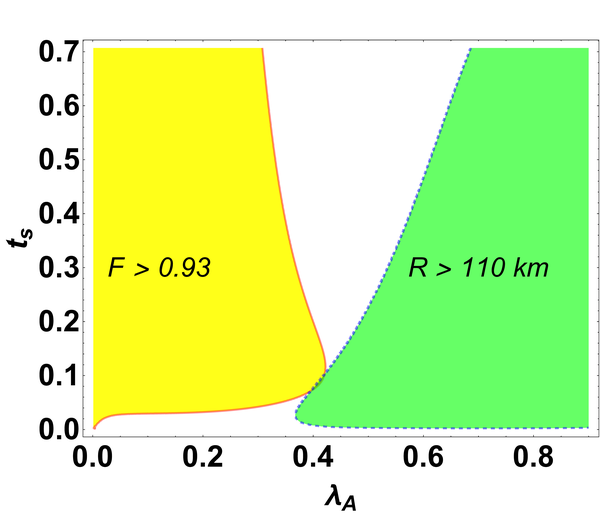}
	\caption{
		\label{fig001NoiseContourRangeFidelity}
		(Color online) The green area shows the regions with the range larger than $110km$. The yellow area shows the regions with the fidelity greater than $0.93$ ($\epsilon = 0.001$). }
\end{figure}
\begin{figure}
	\centering
	\includegraphics[width=0.99\columnwidth]{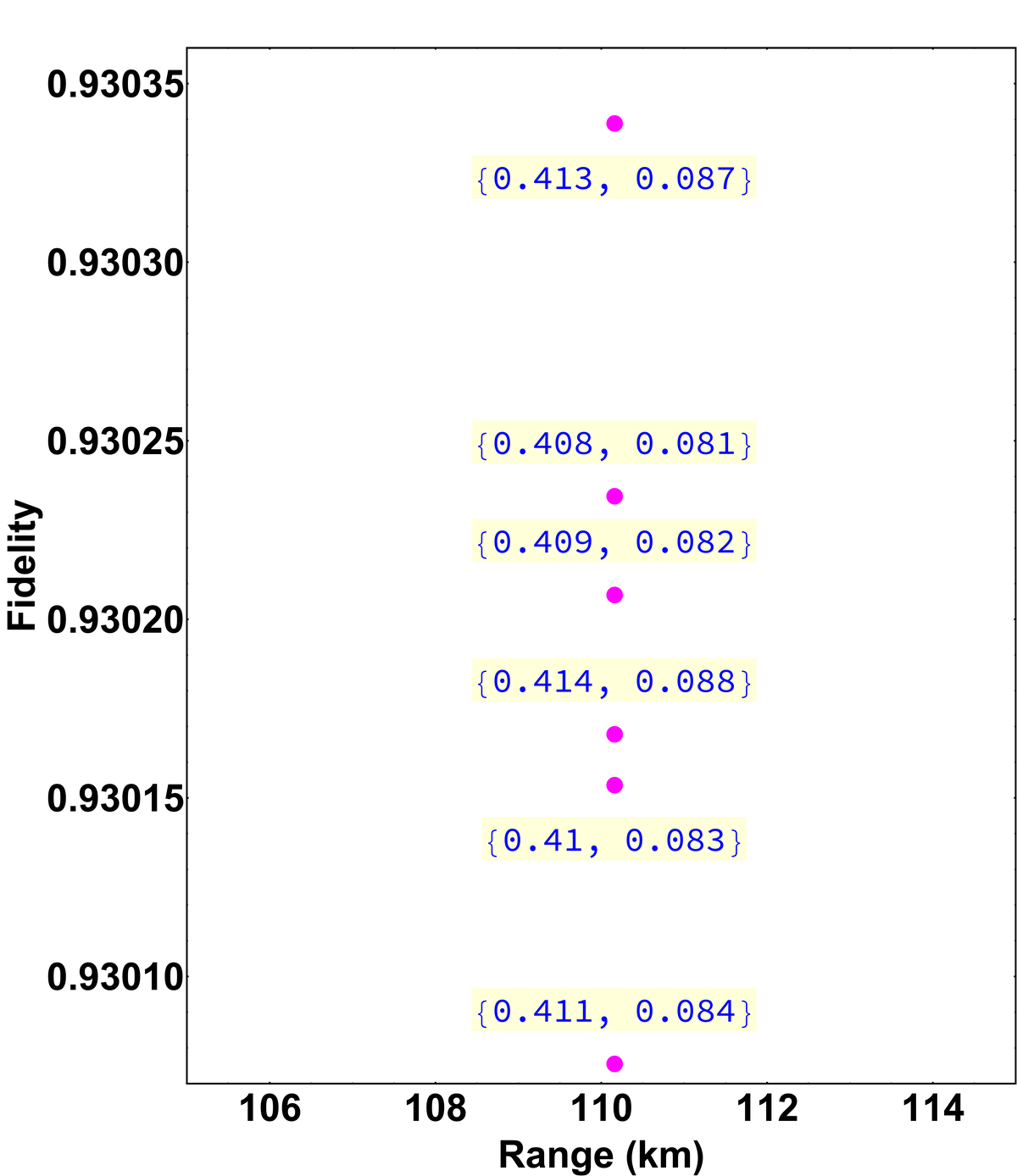}
	\caption{
		\label{fig001NoiseRangeFidelityPoints}
		(Color online) The $(\lambda_A,t_s)$ points with a range larger than $110km$ and a fidelity greater than $0.93$ ($\epsilon = 0.001$). }
\end{figure}
\begin{figure}
	\centering
	\includegraphics[width=0.99\columnwidth]{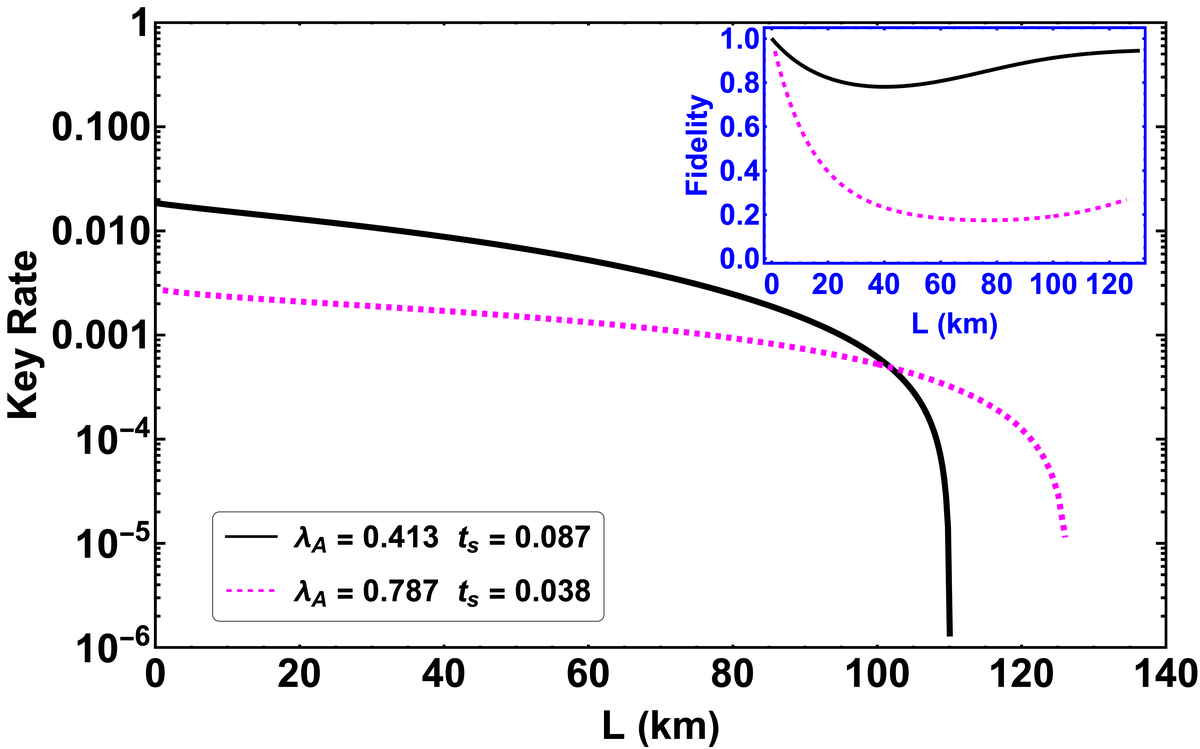}
	\caption{
		\label{fig001NoiseMaxRangeandHighFidelityKeyandFidelity}
		(Color online) Secret key rate (per pulse) versus transmission distance for $\lambda_A=0.787$, $t_s=0.038$ and $\lambda_A=0.413$, $t_s=0.087$, for the case with $\epsilon= 0.001$. The inset compares fidelities. }
\end{figure}
\begin{table}[t!]
	\caption{Range and fidelity of the protocol for different values of the excess noise.} 
	\centering 
	\begin{tabular}{c c c c c c} 
		\hline\hline 
		$\;\;\;\;$ & $\epsilon\simeq 0\;\;\;\;$ & $\epsilon=0.001\;\;\;\;$ & $\epsilon=0.005\;\;\;\;$ & $\epsilon=0.01\;\;\;\;$ & $\epsilon=0.05\;\;\;\;$ \\ [0.5ex] 
		\hline\hline 
		$R (km)$ & 288 & 127 & 92 & 78 & 47 \\ 
		$F$ & 0.36 & 0.27 & 0.26 & 0.23 & 0.25  \\
		$\lambda_A$ & 0.815 & 0.787 & 0.798 & 0.819 & 0.849  \\
		$t_s$ & 0.041 & 0.038 & 0.085 & 0.121 & 0.246  \\
		$\varepsilon_{tm}$ & 0 & 0.346 & 0.341 & 0.353 & 0.385  \\
		$\varepsilon_{rec}$ & 0 & 0.001 & 0.005 & 0.010 & 0.044  \\
		\hline 
		$R (km)$ & 265 & 110 & 76 & 62 & 35 \\ 
		$F$ & 0.94 & 0.93 & 0.93 & 0.93 & 0.92  \\
		$\lambda_A$ & 0.359 & 0.413 & 0.413 & 0.423 & 0.473  \\
		$t_s$ & 0.004 & 0.087 & 0.182 & 0.257 & 0.464  \\
		$\varepsilon_{tm}$ & 0 & 0.157 & 0.161 & 0.164 & 0.201  \\
		$\varepsilon_{rec}$ & 0 & 0.001 & 0.005 & 0.009 & 0.040  \\
		\hline
		$\Delta R (\%)$ & -8 & -13 & -17 & -21 & -26 \\
		$\Delta F (\%)$ & 161 & 244 & 258 & 304 & 268 \\ [1ex] 
		\hline\hline 
	\end{tabular}
	\label{tablenumericalvalues} 
\end{table}

In the intersection of green and yellow regions of Fig.~\ref{figZeroNoiseContourRangeFidelity}, the protocol has proper range and fidelity. The $(\lambda_A,t_s)$ points with $R>265km$ and $F>0.94$ are show in Fig.~\ref{figZeroNoiseRangeFidelityPoints}. In this figure, at a constant value of $t_s$, the fidelity decreases by increasing $\lambda_A$, while the range becomes greater or remains constant by increment of $\lambda_A$. 
These results show that, in the case of $\epsilon \simeq 0$, by fine tuning of the modulation variance of the input state, and the gain of the QS, one can achieve a protocol with a range around $265 km$ and a fidelity of about $0.94$. 

Fig.~\ref{figZeroNoiseMaxRangeandHighFidelityKeyandFidelity} and its inset compare, respectively, the secret key rate and fidelity, for $\lambda_A=0.359$, $t_s=0.004$ (the first case with range $R\simeq265km$ and fidelity $F\simeq0.94$) and $\lambda_A=0.815$, $t_s=0.041$ (the second case with maximum range $R=288km$ and $F=0.36$).
As can be seen, there is a trade-off between protocol range (or key rate) and fidelity. However, it is obvious that although the first case has the range which is about $8\%$ less than the second case, but the fidelity of the first case is approximately $161\%$ larger than the second one. Therefore, a much better fidelity can be obtained at the cost of a slight reduction in protocol range.

Now, we repeat our analysis fo a nonzero value of the excess noise $\epsilon=0.001$. We find that, in this situation, when $\lambda_A=0.787$ and $t_s=0.038$, the maximum range of $R=127km$ is obtained, which is about $56\%$ less than the case with $\epsilon \simeq 0$ (note that the excess noise $\epsilon=0.001$ at distance $L=127km$ is equivalent to $\varepsilon_{tm}\simeq 0.346$ and $\varepsilon_{rec} \simeq 0.001$). In fact, the excess noise reduces the correlation between Alice and Bob modes and brings down the secure transmission distance of the protocol~\cite{jafari2022discrete}. In addition, the fidelity is small and about $F=0.27$. As a result, we search for other values of $(\lambda_A,t_s)$ with considerable range and fidelity. 
In Fig.~\ref{fig001NoiseContourRangeFidelity}, the green and yellow areas correspond to the regions with $R>110km$ and $F>0.93$, respectively.
The general results of this figure are similar to the results of Fig.~\ref{figZeroNoiseContourRangeFidelity}, except that the fidelity is not large at low values of $t_s$. A small value of $t_s$ corresponds to a large value of gain $g$ for QS. Therefore, the reduction of the fidelity for low values of $t_s$ and $\lambda_A$ can be related to the amplification of the noise (instead of the low amplitude signal) at the Bob side, which diminishes the similarity between Alice and Bob states. The $(\lambda_A,t_s)$ points which lie in the intersection of green and yellow regions of Fig.~\ref{fig001NoiseContourRangeFidelity} are shown in Fig.~\ref{fig001NoiseRangeFidelityPoints}, and have acceptable range, $R \simeq  110 km$, and fidelity, $F \simeq 0.93$ (the excess noise $\epsilon=0.001$ at distance $L=110km$ is equivalent to $\varepsilon_{tm}\simeq 0.157$ and $\varepsilon_{rec} \simeq 0.001$). In this case, the range is about $13\%$ less than the maximum achievable range (i.e., 127km) while the fidelity is approximately $244\%$ larger. 
In Fig.~\ref{fig001NoiseMaxRangeandHighFidelityKeyandFidelity} and its inset, we plot the secrete key rate and fidelity versus propagation distance for $\lambda_A=0.413$, $t_s=0.087$ and compare them with the case of optimal range, i.e., $\lambda_A=0.787$, $t_s=0.038$. This figure indicates that both range (not the key rate) and fidelity can not be optimal at the same time, and by decreasing one, the other can be increased.

Finally, Table~\ref{tablenumericalvalues} summarizes the range and fidelity of the protocol, for five different values of the excess noise $\varepsilon$. In this table, the first six rows correspond to the case with maximum achievable range, while the second six rows belong to a case with acceptable range and fidelity. In addition, the last two rows show the variation (in percent) of protocol range and fidelity. Please note that for completeness we also report the equivalent excess noise at the transmitter $\varepsilon_{tm}$ and receiver $\varepsilon_{rec}$ sides (for maximum achievable length $R$). 

This table again confirms that we can obtain much higher fidelity (from $161\%$ to $304\%$), if we give up a smaller amount of the protocol range (from $8\%$ to $26\%$). For all values of the noise, this can be done by reduction of $\lambda_A$, i.e., the modulation variance of the input state. The higher-order number states have
a higher probability of being affected by channel loss. Accordingly, if, by decreasing $\lambda_A$, the weight of higher-order Fock states is reduced, and, at the end of the channel, is properly repaid by a QS (as an amplifier), the fidelity of the protocol can be effectively improved~\cite{jafari2023high}.
Moreover, according to Table~\ref{tablenumericalvalues}, at nonzero values of the excess noise, the amplification gain should be decreased (i.e., $t_s$ should be increased) to obtain higher fidelity. This action can be attributed to the prevention of noise amplification in the system.

\section{Conclusions}
\label{secConclusions}

In summary, by restricting eavesdropper to a Gaussian collective attack, we derived an exact expression for Alice and Bob's states and their fidelity, in a CV-QKD protocol which consists of a ZPC at the Alice side and a QS at the Bob side. For the case of Gaussian modulation, we realized that one can simply remove the ZPC operation and as an alternative reduce the modulation variance (or $\lambda_A$) of the input state. Based on these equations, a direct search algorithm was used to obtain the optimal value of $\lambda_A$ and $t_s$ that maximize the protocol range.
For the case of a pure loss channel ($\epsilon \simeq 0$), the results show that the maximum achievable range is only $6.5\%$ less than the fundamental limit of repeaterless quantum communication. On the other hand, our investigation for both noiseless and noisy channels revealed that, although at the optimal values of $\lambda_A$ and $t_s$, the protocol range is maximum, however, the fidelity is small and less than $0.36$. Therefore, other values of $\lambda_A$ and $t_s$ that give acceptable range and fidelity were obtained. We find that there is a trade-off between range and fidelity, and much higher fidelity can be obtained if someone ignores a smaller amount of the protocol range. This modification can be done by reduction of the modulation variance of the input state, and in the case of a noisy channel, by a decrease in the gain of QS.
\\
\\
%
%
\noindent
{\bf Declaration of Competing Interest  }

The authors declare that they have no known competing financial 
interests or personal relationships that could have appeared to influence 
the work reported in this paper. 
\\
\\
\noindent
{\bf Data availability  }

Data will be made available on request.




\begin{thebibliography}{30}

%
%

\bibitem {wolf2021quantum}
R. Wolf, “Quantum key distribution protocols,” in Quantum Key Distribution, (Springer, 2021).

\bibitem {pirandola2020advances}
S. Pirandola, U. L. Andersen, L. Banchi, M. Berta, D. Bunandar, R. Colbeck, D. Englund, T. Gehring, C. Lupo, C. Ottaviani et al., “Advances in quantum
cryptography,” Adv. optics photonics 12, 1012–1236 (2020).

\bibitem {gisin2002quantum}
N. Gisin, G. Ribordy, W. Tittel, and H. Zbinden, “Quantum cryptography,” Rev. modern physics 74, 145 (2002).

\bibitem {ralph1999continuous}
T. C. Ralph, “Continuous variable quantum cryptography,” Phys. Rev. A 61, 010303 (1999).

\bibitem {PhysRevLett.88.057902}
F. Grosshans and P. Grangier, “Continuous variable quantum cryptography using coherent states,” Phys. Rev. Lett. 88, 057902 (2002).


\bibitem {grosshans2003quantum}
F. Grosshans, G. Van Assche, J. Wenger, R. Brouri, N. J. Cerf, and P. Grangier, “Quantum key distribution using gaussian-modulated coherent states,”
Nature 421, 238–241 (2003).

\bibitem {weedbrook2012gaussian}
C. Weedbrook, S. Pirandola, R. García-Patrón, N. J. Cerf, T. C. Ralph, J. H. Shapiro, and S. Lloyd, “Gaussian quantum information,” Rev. Mod. Phys. 84,
621 (2012).

\bibitem {braunstein2005quantum}
S. L. Braunstein and P. Van Loock, “Quantum information with continuous variables,” Rev. modern physics 77, 513 (2005).

\bibitem {Bennett-2014}
G. Bennett, Charles H.; Brassard, “Quantum cryptography: Public key distribution and coin tossing,” Theor. Comput. Sci. 560, 7–11 (2014).


\bibitem {PhysRevLett.67.661}
A. K. Ekert, “Quantum cryptography based on bell’s theorem,” Phys. Rev. Lett. 67, 661–663 (1991).

\bibitem {bennett1992experimental}
C. H. Bennett, F. Bessette, G. Brassard, L. Salvail, and J. Smolin, “Experimental quantum cryptography,” J. cryptology 5, 3–28 (1992).

\bibitem {hirano2003quantum}
T. Hirano, H. Yamanaka, M. Ashikaga, T. Konishi, and R. Namiki, “Quantum cryptography using pulsed homodyne detection,” Phys. review A 68, 042331
(2003).

\bibitem {yonezawa2007experimental}
H. Yonezawa, S. L. Braunstein, and A. Furusawa, “Experimental demonstration of quantum teleportation of broadband squeezing,” Phys. review letters 99,
110503 (2007).

\bibitem {ma2019long}
H.-X. Ma, P. Huang, D.-Y. Bai, T. Wang, S.-Y. Wang, W.-S. Bao, and G.-H. Zeng, “Long-distance continuous-variable measurement-device-independent
quantum key distribution with discrete modulation,” Phys. Rev. A 99, 022322 (2019).

\bibitem {PhysRevA.102.012608}
L. Hu, M. Al-amri, Z. Liao, and M. S. Zubairy, “Continuous-variable quantum key distribution with non-gaussian operations,” Phys. Rev. A 102, 012608
(2020).

\bibitem {guo2019continuous}
Y. Guo, W. Ye, H. Zhong, and Q. Liao, “Continuous-variable quantum key distribution with non-gaussian quantum catalysis,” Phys. Rev. A 99, 032327
(2019).

\bibitem {ye2019improvement}
W. Ye, H. Zhong, Q. Liao, D. Huang, L. Hu, and Y. Guo, “Improvement of self-referenced continuous-variable quantum key distribution with quantum
photon catalysis,” Opt. express 27, 17186–17198 (2019).

\bibitem {blandino2012improving}
R. Blandino, A. Leverrier, M. Barbieri, J. Etesse, P. Grangier, and R. Tualle-Brouri, “Improving the maximum transmission distance of continuous-variable
quantum key distribution using a noiseless amplifier,” Phys. Rev. A 86, 012327 (2012).

\bibitem {pegg1998optical}
D. T. Pegg, L. S. Phillips, and S. M. Barnett, “Optical state truncation by projection synthesis,” Phys. Rev. Lett. 81, 1604 (1998).

\bibitem {Ralph2009NondeterministicNL}
T. C. Ralph and A. P. Lund, “Nondeterministic noiseless linear amplification of quantum systems,” arXiv: Quantum Phys. 1110, 155–160 (2009).

\bibitem {ghalaii2020long}
M. Ghalaii, C. Ottaviani, R. Kumar, S. Pirandola, and M. Razavi, “Long-distance continuous-variable quantum key distribution with quantum scissors,”
IEEE J. Sel. Top. Quantum Electron. 26, 1–12 (2020).

\bibitem {ghalaii2020discrete}
M. Ghalaii, C. Ottaviani, R. Kumar, S. Pirandola, and M. Razavi, “Discrete-modulation continuous-variable quantum key distribution enhanced by quantum
scissors,” IEEE J. on Sel. Areas Commun. 38, 506–516 (2020).

\bibitem {bartley2015directly}
T. J. Bartley and I. A. Walmsley, “Directly comparing entanglement-enhancing non-gaussian operations,” New J. Phys. 17, 023038 (2015).


\bibitem{matsumoto2021survey}
R. Matsumoto and M. Hagiwara, “A survey of quantum error correction,” IEICE Trans. on Fundam. Electron. Commun. Comput. Sci. 104, 1654–1664
(2021).

\bibitem{salim2020enhancing}
 S. I. Salim, A. Quaium, S. Chellappan, and A. A. Al Islam, “Enhancing fidelity of quantum cryptography using maximally entangled qubits,” in GLOBECOM
2020-2020 IEEE Global Communications Conference, (IEEE, 2020), pp. 1–6.

\bibitem{bouchard2022quantum}
F. Bouchard, D. England, P. J. Bustard, K. Heshami, and B. Sussman, “Quantum communication with ultrafast time-bin qubits,” PRX Quantum 3, 010332
(2022).

\bibitem{PhysRevLett.108.130503}
H.-K. Lo, M. Curty, and B. Qi, “Measurement-device-independent quantum key distribution,” Phys. Rev. Lett. 108(13), 130503 (2012).

\bibitem{PhysRevA.89.052301}
Z. Li, Y.-C. Zhang, F. Xu, X. Peng, and H. Guo, “Continuous-variable measurement-device-independent quantum key distribution,” Phys. Rev. A 89(5), 052301 (2014).

\bibitem{PhysRevA.87.022308}
C. Weedbrook, “Continuous-variable quantum key distribution with entanglement in the middle,” Phys. Rev. A 87(2),
022308 (2013).

\bibitem{HighRate397NatPhotonics}
S. Pirandola, C. Ottaviani, G. Spedalieri, C. Weedbrook, S. Braunstein, S. Lloyd, T. Gehring, C. Jacobsen, and U. Andersen, “High-rate measurement-device-independent quantum cryptography,” Nat. Photonics 9(6), 397–402 (2015).

\bibitem {jafari2023high}
K. Jafari, M. Golshani, and A. Bahrampour, “High-fidelity continuous-variable quantum key distribution via the proper combination of zero-photon catalysis
and quantum scissors,” JOSA B 40, 661–666 (2023).

\bibitem {jafari2022discrete}
K. Jafari, M. Golshani, and A. Bahrampour, “Discrete-modulation measurement-device-independent continuous-variable quantum key distribution with a
quantum scissor: exact non-gaussian calculation.” Opt. Express 30, 11400–11423 (2022).

\bibitem {leverrier2009unconditional}
A. Leverrier and P. Grangier, “Unconditional security proof of long-distance continuous-variable quantum key distribution with discrete modulation,” Phys.
review letters 102, 180504 (2009).

\bibitem {djordjevic2019physical}
I. B. Djordjevic, Physical-layer security and quantum key distribution (Springer, 2019).

\bibitem {mivcuda2012noiseless}
M. Mi{\v{c}}uda, I. Straka, M.Mikov{\'a}, M. Du{\v{s}}ek, N. J. Cerf, J. Fiur{\'a}{\v{s}}ek, and M. Je{\v{z}}ek, “Noiseless loss suppression in quantum optical communication,” Phys. Rev. Lett. 109, 180503 (2012).

\bibitem {arfken2011mathematical}
G. B. Arfken, H. J. Weber, and F. E. Harris, Mathematical methods for physicists: a comprehensive guide (Academic press, 2011).

\bibitem {winnel2021overcoming}
M. S. Winnel, J. J. Guanzon, N. Hosseinidehaj, and T. C. Ralph, “Overcoming the repeaterless bound in continuous-variable quantum communication
without quantum memories,” arXiv preprint arXiv:2105.03586 (2021).

\bibitem {pirandola2017fundamental}
S. Pirandola, R. Laurenza, C. Ottaviani, and L. Banchi, “Fundamental limits of repeaterless quantum communications,” Nat. communications 8, 15043
(2017).
\end{thebibliography}

%
%
%
%
%

\end{document}